\documentclass[a4paper,12pt]{article}
\usepackage[warn]{mathtext}				
\usepackage[T1, T2A]{fontenc}			
\usepackage[utf8]{inputenc} 			
\usepackage[warn]{mathtext}			 
\usepackage[T1, T2A]{fontenc}			
\usepackage[utf8]{inputenc} 			 
\usepackage{wrapfig}				
\usepackage{amssymb, amsmath}			 
\usepackage{indentfirst}   
\usepackage[titletoc]{appendix}
\usepackage{mathrsfs}
\usepackage{cmap} 			
\usepackage[pdftex]{graphicx}
\usepackage{pgfplotstable}	
\usepackage[bottom]{footmisc}
\usepackage[hidelinks]{hyperref}
\DeclareGraphicsExtensions{.eps}
\newcommand*{\defeq}{\stackrel{\text{def}}{=}}
\usepackage{graphicx}

\usepackage{float}
\usepackage[left=20mm,right=20mm,top=20mm,bottom=20mm,bindingoffset=0mm]{geometry}
\let\DS     = \displaystyle

\def\edited#1{{#1}}

\iffalse

\else

\title{Unsteady thermal transport in \edited{an} instantly heated semi-infinite free end Hooke chain}

\author{Sergei D. Liazhkov$^{1,2}$}
\date{%
    $^1$Peter the Great Saint-Petersburg Polytechnic University, St. Petersburg, Russia\\%
    $^2$Institute for Problems in Mechanical Engineering of the Russian Academy of Sciences, St. Petersburg, Russia.\\[4mm]%
    \today
}

\begin{document}

\maketitle

\abstract{We consider unsteady ballistic heat transport in a semi-infinite Hooke chain with free end and \edited{arbitrary} initial temperature profile.  An analytical description of the evolution of the kinetic temperature is proposed in both discrete~(exact) and continuum~\edited{(approximate)} formulations. By comparison of \edited{the} discrete and continuum descriptions \edited{of} kinetic temperature field, we reveal some restrictions to the latter. Specifically, the far-field kinetic temperature is well described by the continuum solution, which, however, \edited{deviates} near and at the free end~(boundary). We show analytically that, \edited{after} thermal wave reflects from the boundary, the discrete solution for the kinetic temperature undergoes a jump near the free end. A comparison of the descriptions of heat propagation in the semi-infinite and infinite Hooke chains is presented. \edited{Results of the current paper are expected to provide insight into non-stationary heat transport in the semi-infinite lattices. }}





\maketitle
\section{Introduction}\label{sec1}
 Heat transfer at \edited{the} macroscale is known to be diffusive and to obey the Fourier law. Using of the law as \edited{a} constitutive relation, a wide class of problems in continuum mechanics can be solved~(see, e.g.,~\cite{Rick}). However, theoretical studies~\cite{Rieder, Lepri, Dhar2008} and experiments~\cite{Chang,Graphite,Maznev,Anufriev, Anufriev2, Chang2,Xu} show that at \edited{the} micro- and nanoscale heat propagation is nondiffusive, e.g., ballistic. In particular, \edited{deviations} from the Fourier law \edited{are} demonstrated in nanotubes~\cite{Chang}, silicon membranes~\cite{Maznev}, silicon nanowires~\cite{Anufriev, Anufriev2}, graphene~\cite{Xu}. Therefore, building of theories, describing thermal processes at \edited{the} micro- and nanoscale, is required and is relevant also for the reason of innovative development of micro- and nanoelectronics~\edited{(see, e.g.,~\cite{Colloq, Shahil, Malik, HAMR, Moore})}. 
    
    To the best of our knowledge, two approaches are used in general for \edited{the} analytical description of heat transfer at \edited{the} microscale (or nanoscale), namely the lattice dynamics (LD) approach and \edited{the} kinetic theory. By dint of investigation of the Boltzmann transport equation~(BTE), one can solve problems, which are unsolvable by the LD method~(see, e.g.,~\cite{Majumdar, Cahil, Spohn}). Since the BTE is continuum, quantities, obtained through BTE, change in space also continuously. Therefore, there may be some restrictions \edited{on} description of heat transport at \edited{the} nanoscale, where one may have necessity to deal with discrete structures. Hence a question arises what these restrictions are. 
    
    In paper~\cite{Kuz2021}, the kinetic theory of unsteady heat transport in the infinite one-dimensional harmonic chain is linked with the LD theory. The \edited{approximate solution} for the kinetic temperature is derived in the continnum limit using the discrete\footnote{i.e. changing in dependence of the particle number.}~(exact) solution, obtained by the LD approach. It is shown \edited{that} the result is in a practically ideal agreement with the kinetic theory. Nonetheless, it remains still unclear, whether the continuum solution, obtained by one or another method, corresponds to the exact one \edited{always}.
    
    In general, one can distinguish discrete and continuum analytical descriptions of the ballistic heat transport. In the pioneering work of Klein and \edited{Ilya} Prigogine~\cite{Klein}, the evolution law of the discrete field of the kinetic temperature was obtained using exact Schr{\"o}dinger solution~\cite{Schroedinger}\footnote{See also English translation of the Schr{\"o}dinger article~\cite{Schroedinger_renew}.} of the \edited{dynamical} equations for the Hooke chain\footnote{\edited{This is the monoatomic harmonic chain of identical particles, connected by the linear identical springs, see~\cite{ZAMM}.}}. \edited{The result was reproduced by Hemmer in his PhD thesis~\cite{Hemmer}.} In the pioneering work by Krivtsov~\cite{Krivtsov2015}, the PDE for the continuum field of the kinetic temperature was derived, solution of which is proposed in the integral form. In papers~\cite{Sokolov2021,Gavr2022}, the discrete and continuum descriptions of kinetic temperature fields are compared. It is shown in~\cite{Sokolov2021} that \edited{the} evolution over time of the temperature fields caused by arbitrary initial perturbation~(except point) leads to coincidence of these. However, aforesaid results are about energy propagation in \edited{the}~\textit{infinite} chains only. 
         
    The question of heat transport in the finite or semi-infinite lattices remains open. Some analytical treatments to describe it are proposed in studies~\cite{Guzev, Gudim}. In paper~\cite{Guzev}, \edited{the solution} for the kinetic temperature in the finite\footnote{Ends of chain are connected with the fixed points by linear stiffness springs.} Hooke chain was obtained. Gudimenko~\cite{Gudim} found approximate solution of heat transfer problem in \edited{the} semi-infinite chain with an absorbing boundary. Despite the obtained results, many ambiguities remain, for instance, behavior of quantities at the boundary or influence of other boundary conditions on heat propagation. In particular, answering aforementioned problems, related to free boundaries, is necessary for development of theoretical models of the experiments associated, \edited{e.g.}, with reflection of phonons~\edited{\cite{Wolfe, Northrop, Ravi}}.
    
    In this paper, we describe the process of heat transport in the semi-infinite free end Hooke chain. Firstly, we derive \edited{the} exact \edited{solution} for the kinetic temperature, and, following~\cite{Kuz2021}, we perform \edited{an approximation} of it in the continuum limit. Then, analogously to~\cite{Sokolov2021, Gavr2022}, we compare \edited{the} discrete~(exact) and continuum descriptions of the heat transport. \edited{By} the comparison, we reveal discrepancies between these descriptions near and \edited{at} the boundary.  
    
    The paper is organized as follows. In~Sect.\ref{sect2}, we formulate the problem and derive exact expression for particle velocities~\edited{(Sect.\ref{Sect21})}, which is further applied to derive the exact expression for the kinetic temperature~\edited{(Sect.\ref{Sect22})}. In~Sect.\ref{sect3}, we determine the kinetic temperature in the continuum limit.  In~Sect.\ref{sect4}, the fundamental solution for the kinetic temperature in the continuum limit is derived, which is present as \edited{a} sum of the contributions from incident and reflected thermal waves. We reveal \edited{an} interrelation between the continuum solutions \edited{for the kinetic temperature in} the semi-infinite and infinite Hooke chains.  In~Sect.\ref{sect5}, the discrete and continuum solutions for kinetic temperature fields are compared. Examples of the rectangular~(Sect.\ref{subsect52}) and step~(Sect.\ref{subsect53}) initial perturbations are considered. In Sect.\ref{sect6}, we compare \edited{the} theory of ballistic heat transport in the semi-infinite and infinite Hooke chains and find \edited{an} interrelation between the corresponding \edited{discrete} solutions for the kinetic temperature. In Sect.\ref{sect7}, results of the paper are discussed.

\section{Discrete solution for the kinetic temperature}\label{sect2}
\subsection{Formulation of the problem and derivation of expression for particle velocities}\label{Sect21}

We consider the semi-infinite Hooke chain\footnote{\edited{Definition of the Hooke chain is given in Sect.\ref{sec1} and in~\cite{ZAMM}.}}, having one free end and assume that the particles of the chain interact with the nearest neighbors. Therefore, the \edited{dynamical} equations can be written as
\begin{equation}
    \begin{array}{l}
    \displaystyle \dot{u}_n=v_n,\\[1mm]
    \displaystyle \dot{v}_0=\omega_e^2(u_1-u_0),\\
     \displaystyle\dot{v}_n=\omega_e^2(u_{n+1}-2u_n+u_{n-1}),\quad n\in \mathbb{N},\quad \omega_e=\sqrt{c/m}, \label{DYNeq}
    \end{array}
\end{equation}
where~$m$ is the particle mass;~$c$ is the \edited{spring stiffness};~$u_n$ and~$v_n$ are displacement and velocity of particle~$n$ respectively. The equations are supplemented by the following initial conditions:
\begin{equation}
    u_n=0,\edited{\quad v_n=\mathcal{V}_n}. \label{IC}
\end{equation}
\edited{Here~$\mathcal{V}_n$ is the initial velocity field such that}
\begin{equation}
   \mathcal{V}_n=\rho_n\sqrt{\frac{k_\mathrm{B}T_n^0}{m}}, \label{defT0}
\end{equation}
where~$k_{\mathrm{B}}$ is the Boltzmann constant;~$T_n^0$ is the initial kinetic temperature of particle~(see definition~(\ref{DefT}));~$\rho_n$ are uncorrelated random numbers with zero mean and unit variance:
\begin{equation}
   \langle\rho_n\rangle=0, \qquad \langle \rho_{j}\rho_{n}\rangle = \delta_{jn}, \label{UNCS}
\end{equation}
where~$\delta_{jn}$ is the Kronecker delta; ~$\langle...\rangle$ stands for the mathematical expectation sign. Therefore, the initial conditions~(\ref{IC}) \edited{with~(\ref{defT0})} and~(\ref{UNCS}) imply an existence of some initial temperature field in the chain and zero initial heat fluxes\footnote{The statement of problem corresponds to experimental heating of the crystal by the ultrashort laser pulse. Since the \edited{expression} for heat flux in the Hooke chain contains covariances of displacements and velocities~(see, e.g.,~\cite{Lepri, Krivtsov2018}), then initial zero field of initial displacements means zero initial heat fluxes.}~\cite{Krivtsov2015}.

In order to define \edited{the} kinetic temperature \edited{in} the chain, we introduce \edited{an} infinite set of realizations with different initial conditions~(\ref{IC}). For \edited{the} one-dimensional Hooke chain, we determine the kinetic temperature,~$T_n$, as follows:\footnote{We determine the kinetic temperature by its statistical definition~(see, e.g., chapter 3, Sect.~29 in~\cite{Landau}). Unambiguous definition of the temperature for systems far from equilibrium is still unresolved fundamental problem~(see, e.g.,~\edited{\cite{Casas_Jou,Piglis}}). In this paper, we calculate the kinetic temperature as average of kinetic energy over realizations, because it has simple physical meaning. Discussion of the \edited{ergodicity} remains out of frameworks of this study.}

\begin{equation}
   m\langle v_n^2 \rangle \defeq k_\mathrm{B} T_n. \label{DefT}
   \end{equation}
In general, two approaches are followed to obtain the kinetic temperature in the harmonic crystals. The first involves introducing of covariances of particle displacements and velocities and transformation of the stochastic \edited{dynamical} equations to the deterministic PDE with respect to these covariances. This \edited{approach} is extensively studied in e.g.,~\cite{Krivtsov2015, Krivtsov2014, Kuz2017, Gavr2018}. The second approach is to substitute exact expression for particle velocity into~(\ref{DefT}). One possible way to solve the equations~(\ref{DYNeq}) analytically is to reformulate the \edited{dynamical} problem for the finite chain with two free ends, exact solution of which is known~\cite{Hemmer, Takizawa} and then to proceed to the thermodynamic limit. \edited{The second way, based on operating of the difference equations, is described in~\cite{Lee1, Lee2}.  However, these ways are harder than one, which is proposed below.}

We introduce \edited{the} direct and inverse discrete cosine transforms~(DCT) as follows~\cite{AHMED}
\begin{equation}\label{DCT}
    \DS \hat{u}(\theta)=\sum_{n=0}^\infty u_n\cos \frac{(2n+1)\theta}{2},\quad u_n=\frac{1}{\pi}\int_{-\pi}^\pi \hat{u}(\theta)\cos \frac{(2n+1)\theta}{2}\mathrm{d}\theta,
\end{equation}
where~$\theta$ is the wave number; $\hat{u}$ is some time-dependent function. Note that representation for the particle displacement~(\ref{DCT}) satisfies free boundary condition.
Applying~DCT~(\ref{DCT}) to Eqs.~(\ref{DYNeq}---\ref{IC}) yields equation
\begin{equation}\label{IM_ODE}
   \ddot{\hat{u}}+\omega^2 \hat{u}=0,\quad \edited{\omega(\theta)=2\omega_e \Big\vert\sin \frac{\theta}{2}\Big\vert}, 
\end{equation}
where~$\displaystyle\omega$ is~\edited{the} dispersion relation for the Hooke chain, with~\edited{the} initial conditions 
\begin{equation}\label{IM_IC}
    \hat{u}=0,\quad \dot{\hat{u}}=\sum_{n=0}^{\infty}\edited{\mathcal{V}_n} \cos{\frac{(2n+1)\theta }{2}},
\end{equation}
whence 
\begin{equation}\label{IM_ODE_SOL}
    \hat{u}=\frac{\sin{(\omega t)}}{\omega}\sum_{n=0}^{\infty}\edited{\mathcal{V}_n} \cos{\frac{(2n+1)\theta}{2}}.
\end{equation}
Note that the dispersion relations for the semi-infinite and infinite chains coincide. Applying \edited{the} inverse DCT to~(\ref{IM_ODE_SOL}) with subsequent differentiation with respect to time gives the following \edited{Eq.} for the particle velocity:
\begin{equation}\label{Vel_expr2}
    v_n=\frac{1}{\pi}\sum_{j=0}^{\infty}\edited{\mathcal{V}_j}\int_{-\pi}^{\pi}\cos{\frac{(2j+1)\theta}{2}}\cos{\frac{(2n+1)\theta}{2}}\cos({\omega (\theta)t})\mathrm{d}\theta.
\end{equation}
    Thus, we have the exact expression for velocity of each particle in the semi-infinite chain. In the next subsection, the \edited{Eq.}~(\ref{Vel_expr2}) is employed to obtain the kinetic temperature.

\subsection{Exact expression for the kinetic temperature}\label{Sect22}

Substitution of the solution for particle velocity~(\ref{Vel_expr2}) to~(\ref{DefT}) using uncorrelatedness of the initial velocity field~(\ref{UNCS}) \edited{and~(\ref{defT0})} yields 
\begin{equation}\label{DisT2}
\begin{array}{l}
   \DS T_n=\frac{1}{\pi^2}\sum_{j=0}^\infty T_j^0 \Biggl(\int_{-\pi}^{\pi}\cos{\frac{(2j+1)\theta}{2}}\cos{\frac{(2n+1)\theta}{2}\cos({\omega(\theta) t})} \mathrm{d}\theta\Biggr)^2.
    \end{array}
\end{equation}
The \edited{Eq.}~(\ref{DisT2}) is the exact solution for the kinetic temperature in the semi-infinite chain with free end and will be further referred to as \edited{the}~\textit{discrete solution}.  \edited{Recall} that the discrete solution for kinetic temperature in the \textit{infinite} Hooke chain has \edited{the} form~\cite{Klein, Sokolov2021, Hemmer}
\begin{equation}\label{DisT}
    {T_\mathrm{inf}}_n=\sum_{j=-\infty}^\infty T_j^0 J_{2(n-j)}^2(2\omega_e t),
\end{equation}
where~$J$ is the Bessel function of the first kind. From comparison of Eqs.~(\ref{DisT2}) and~(\ref{DisT}) \edited{it follows that equation for the discrete solution for kinetic temperature in the Hooke chain with arbitrary boundary conditions can be constructed as}
\begin{equation}\label{conv}
\begin{array}{l}
\DS \edited{T_n=\sum_{j \in \mathbb{P}} T_j^0 \left(\frac{\dot{\Phi}_{nj}(t)}{\omega_e}\right)^2}, 
\end{array}
\end{equation}
where~\edited{$\Phi_{nj}(t)$} is the solution of the equation\edited{
\begin{equation}\label{fundPhi}
\begin{array}{l}
\DS \ddot{\Phi}_n-\omega_e^2\mathcal{L}_n\Phi_n=\omega_e \delta(t)\delta_{nj}, \quad n\in\mathbb{P},\\[2mm]
\end{array}
\end{equation}
where~$\mathcal{L}_n$ is the linear difference operator, which depends on specific boundary conditions;}~$\delta(t)$ is the Dirac delta function;~$\mathbb{P}$ is set of numbers, by which particles in the system are indexed. The function~\edited{$\Phi_n$} is supplemented by the initial condition~\cite{Vladimirov}:
\begin{equation}
    \Phi_n \vert_{t<0}=0.      \label{fundPhi2}
\end{equation}
The \edited{Eq.}~(\ref{DisT2}) is further employed to obtain the kinetic temperature in the continuum limit.

\section{Kinetic temperature in the continuum limit}\label{sect3}

In the section, we derive the kinetic temperature in the continuum limit, namely as a function of the continuum coordinate,~$x$. This  representation is suitable for general case when \edited{the} expression~(\ref{DisT2}) becomes hard to use. We show that kinetic temperature in the continuum limit can be expressed as
\begin{equation}\label{ContT}
    \edited{ \begin{array}{l}
   \DS  T(x,t)=T^F(x,t)+T^S(x,t),\\[2mm]
    \DS T^F=\frac{T^0(x)}{2}J_0(4\omega_et),\quad \DS T^S=\frac{1}{2\pi}\int_{0}^\pi T^0(\vert x+v_st\cos{\theta}\vert)\mathrm{d}\theta,
    \end{array}}
\end{equation}
where~$v_s \edited{\defeq} \omega_ea$ is the speed of sound; $a$ is the equilibrium distance~(length of undeformed bond between particles); \edited{$T^0(x)$ is the continuum field of the kinetic temperature such that $T^0(an)=T_n^0$~(see derivation for details).} We further refer~(\ref{ContT}) to as~\edited{the} \textit{continuum solution}. Derivation of~\edited{Eq.(\ref{ContT})} is given below. 

\subsection{Continualization}\label{subsect30}
We use an approach, proposed in paper~\cite{Kuz2021}. First of all, we separate \edited{Eq}.~(\ref{DisT2}) into two terms, corresponding to the two physical processes:
\begin{equation}\label{Divid}
\small
\begin{array}{l}
\DS T_n=T_n^F+T_n^S,\\[2mm]
\DS T_n^F=\sum_{j=0}^\infty T_j^0 F_{nj},\quad T_n^S=\sum_{j=0}^\infty T_j^0 S_{nj},\\[2mm]
\DS F_{nj}=\frac{1}{2\pi^2}\iint_{-\pi}^{\pi}\cos{\frac{(2j+1)\theta_1}{2}}\cos{\frac{(2j+1) \theta_2}{2}}\cos{\frac{(2n+1)\theta_1}{2}}\times \\
\DS \qquad \qquad \qquad \qquad \qquad \qquad \times\cos{\frac{(2n+1)\theta_2}{2}}\cos \left({(\omega(\theta_1)+\omega(\theta_2)) t}\right) \mathrm{d}\theta_1\mathrm{d}\theta_2,\\[3mm]
\DS S_{nj}= \frac{1}{2\pi^2}\iint_{-\pi}^{\pi} \cos{\frac{(2j+1)\theta_1}{2}}\cos{\frac{(2j+1) \theta_2}{2}}\cos{\frac{(2n+1)\theta_1}{2}}\times \\
\DS \qquad \qquad \qquad \qquad \qquad \qquad \times\cos{\frac{(2n+1)\theta_2}{2}}\cos \left({(\omega(\theta_1)-\omega(\theta_2)) t}\right) \mathrm{d}\theta_1\mathrm{d}\theta_2.
\end{array}
\end{equation}
We further show that the first term corresponds to high-frequency oscillations of the kinetic temperature, caused by equilibration of the kinetic and potential energies~\cite{Krivtsov2014}. This is a fast process, occurring in time of order of several hundreds atomic periods. The second term corresponds to the slow process caused by ballistic heat transport.  Characteristic time scale of this process is \edited{much} larger than \edited{one} of the thermal equilibration. 
Following~\cite{Kuz2021}, we perform a continualization for the~\textit{slow} and~\textit{fast} terms of the kinetic temperature,~$T^S$ and~$T^F$ respectively.

\begin{figure}[htb]
\center{\includegraphics[width=0.55\linewidth]{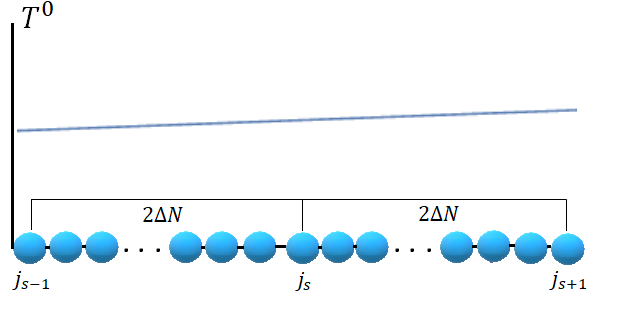}}
\caption{Initial temperature profile.}
\label{fig1}
\end{figure}

We introduce a mesoscale, which is larger than the distance between particles,~$a$, but smaller than macroscale,\footnote{A macroscale can be interpreted as a scale of the order of the length of chain.}~$\mathcal{A}$ and divide the chain into the equal intervals, indexed by~$s$. Each interval~$s$ has the length~$2a\Delta N,\;~\Delta N\gg 1,\; a\Delta N \ll \mathcal{A}$ and is limited by the boundary-particles,~$j_s$. We assume that the initial temperature profile, enclosed in the intervals~$s$, changes slowly~(see Fig.\ref{fig1}).
Therefore, the length of the mesoscale is~$2a\Delta N$. Then, the expression for~$T^S_n$ in~(\ref{Divid}) can be rewritten as
\begin{equation}\label{LimREP}
\begin{array}{l}
    \DS T_n^{S}=\sum_{s=0}^\infty \sum_{j=j_s-\Delta N+1}^{j_s+\Delta N} T_j^0 \;{S}_{nj}\approx \sum_{s=0}^\infty T_{j_s}^0 \sum_{j=j_s-\Delta N+1}^{j_s+\Delta N}{S}_{nj}\\[2mm]
    
    \DS =2a\Delta N \sum_{s=0}^\infty T_{j_s}^0 g_{n,j_s}^{S}(\Delta N),\quad g_{n,j_s}^{S}(\Delta N)=\frac{1}{2a\Delta N} \sum_{j=j_s-\Delta N+1}^{j_s+\Delta N} {S}_{nj},
    \end{array}
\end{equation}
where~$g_{n,j_s}(\Delta N)$ determines contribution of the point~$j_s$ to the \edited{kinetic} temperature at point~$n$. The expression~(\ref{LimREP}) is also valid for~$T^F$ if we replace~$S$ by~$F$. It can also be interpreted as a discrete analogue of \edited{the} fundamental solution, obtained by averaging of~$T_n$ over the mesoscale. Further, we derive the continuum fundamental solutions, corresponding to the slow term,~$g^S$~(see~Sect.\ref{subsect31}) and to the fast term,~$g^F$~(see~Sect.\ref{subsect32}). 

\subsection{Slow term}\label{subsect31}
The following \edited{expression} for~$g_{n,j_s}^S(\Delta N)$ is calculated up to the order~$O\left(\frac{1}{\Delta N}\right)$~(see Appendix~A) and is represented below:
\small
\begin{equation}\label{DiscFUND}
     \begin{array}{l}
     
      \edited{\DS  g_{n,j_s}^S(\Delta N)\approx\frac{1}{16\pi^2 a}\iint_{-\pi}^{\pi}\Bigg[\cos{(\Delta\theta \omega'(\theta_1)t)} \Big[\cos({(n+j_s)\Delta \theta})+\cos({(n-j_s)\Delta \theta})}\\[2mm] 
 \edited{+\DS \cos{(\theta_1(2n+1)-(n+j_s)\Delta \theta)}+\cos({\theta_1(2n+1)-(n-j_s)\Delta \theta})\Big]\mathrm{sinc}(\Delta N \Delta \theta)\Bigg]\mathrm{d}\theta_1\mathrm{d}\theta_2,}\\[2mm]
     \end{array}
\end{equation}
 \normalsize
where~$\Delta \theta=\theta_1-\theta_2$, $()'=\frac{\mathrm{d}}{\mathrm{d}\theta_1}$, $\DS\mathrm{sinc}(x)=\frac{\sin{x}}{x}$.
We change the variables~$\theta=\theta_1$, $q=\Delta \theta$ and rewrite~(\ref{DiscFUND}), using trigonometric identities and symmetry of the integrands with respect to zero: 
    \begin{equation}\label{wpcgs}
    \begin{array}{l}
    \edited{\DS g_{n,j_s}^S(\Delta N) \approx \frac{1}{4\pi}\int_0^\pi \Bigg[ \cos^2\frac{(2n+1)\theta}{2}\Big[\psi_1(n+j_s+\omega'(\theta)t)+\psi_1(n+j_s-\omega'(\theta)t)}\\[2.5mm]
    \edited{+\DS \psi_1(n-j_s+\omega'(\theta)t)+\psi_1(n-j_s-\omega'(\theta)t)\Big]}\\[3.5mm]
    \edited{+\DS \frac{\sin{((2n+1)\theta)}}{2}\Big[\psi_2(n+j_s+\omega'(\theta)t)+\psi_2(n+j_s-\omega'(\theta)t)}\\[3.5mm]
    \edited{+\psi_2(n-j_s+\omega'(\theta)t)+\psi_2(n-j_s-\omega'(\theta)t)\Big]\Bigg]\mathrm{d}\theta,}\\[3.5mm]
   \DS \psi_{1}(\Xi)=\frac{1}{2\pi a}\int_{\theta-\pi}^{\theta+\pi}\cos{(\Xi q)}\mathrm{sinc}(q \Delta N) \mathrm{d}q,\\[3.5mm]
   \DS \psi_{2}(\Xi)=\frac{1}{2\pi a}\int_{\theta-\pi}^{\theta+\pi}\sin{(\Xi q)}\mathrm{sinc}(q \Delta N) \mathrm{d}q,
    \end{array}
    \end{equation}
 where~$\psi_1$ and~$\psi_2$ are referred to as wave packets propagating with group velocity~$v_g=a\omega'$. Averaging of the function~$g_{n,j_s}^S$ over \edited{the} mesoscale leads to a sum of the integrals of these wave packets. In the limit case~($\Delta N \gg 1)$ the wave packet~$\psi_2$ is negligible and the expression for~$\psi_1$ has the following approximate form~(see Appendix~B):
 \begin{equation}\label{PSI1}
    \begin{array}{l}
    \psi_1(\Xi)\approx \DS\frac{1}{2a\Delta N}H\left(1-\frac{\vert \Xi \vert}{\Delta N}\right),
    \end{array}
\end{equation}
where~$H(x)$ is the Heaviside function.  \edited{Taking into account~$\Delta N \gg 1$}, we obtain the final form of the discrete fundamental solution:
   \begin{equation}\label{DiscFUND2}
    \begin{array}{l}
     \edited{\DS   g_{n,j_s}^S(\Delta N)\approx\frac{1}{4\pi}\int_0^\pi \Bigg[\psi(n-j_s+\omega'(\theta)t, \Delta N)+\psi(n-j_s-\omega'(\theta)t, \Delta N)}\\[3mm]
     \edited{\DS+\psi(n+j_s+\omega'(\theta)t, \Delta N)+\psi(n+j_s-\omega'(\theta)t, \Delta N)\Bigg]\mathrm{d}\theta},\\[3.5mm]
     \DS \psi(\Xi,\Delta N)=\frac{1}{2a\Delta N}\cos^2{\left(\DS\frac{\theta(2n+1)}{2}\right)}H\left(1-\frac{\vert \Xi \vert}{\Delta N}\right).
        \end{array}
    \end{equation}
    We introduce continuous functions~$T^S(x),\; T^0(x),\; g_c^S(x,y)$ such that
    \begin{equation}\label{LimREP2}
    \begin{array}{l}
       \DS T^0(an)=T_n^0,\quad g_c^S(x,y)=\lim_{{\frac{a\Delta N}{\mathcal{A}}\rightarrow 0}}g_{n,j_s}^S(\Delta N),\\[2mm]
      \DS T^S(x)=\lim_{{\frac{a\Delta N}{\mathcal{A}}\rightarrow 0}}T^S_n=\int_0^\infty T^0(y)g_c^S(x,y)\mathrm{d}y.
        \end{array}
    \end{equation}
 In the limit case~$a\Delta N/\mathcal{A}\rightarrow 0$, the function~$\psi$ can be replaced by the Dirac delta function. Using \edited{Eqs.}~($\ref{DiscFUND2}$) and~($\ref{LimREP2}$), we obtain the fundamental solution for the slow term of the kinetic temperature:
    \begin{equation}\label{ContFUND}
    \begin{array}{l}
       \DS g_c^S(x,y)=g^S(x-y)+g^S(x+y),\\ \edited{\DS \quad g^S(x)=\frac{1}{4\pi}\int_0^\pi \DS \Big[\delta(x+v_g(\theta)t)+\delta(x-v_g(\theta)t)\Big]\mathrm{d}\theta},
        \end{array}
    \end{equation}
where~\edited{$v_g(\theta)=\DS v_s  \cos{\frac{\theta}{2}} \,\mathrm{sgn}\left(\sin \frac{\theta}{2}\right) $} is the group velocity.~\footnote{Here evenness \edited{property} of the Dirac delta function \edited{is} used.}

The slow term~$T^S$ is further obtained as the integral convolution of this fundamental solution with field of the initial temperature:

\begin{equation}\label{Slow1}
\small
\edited{\begin{array}{l}
   \DS T^S=\frac{1}{4\pi}\int_0^\infty T^0(y)\Bigg[\int_0^\pi \delta(x-y+v_g(\theta)t)\,\mathrm{d}\theta+\int_0^\pi \delta(x-y-v_g(\theta)t)\,\mathrm{d}\theta \\[2mm]+\DS\int_0^\pi \delta(x+y-v_g(\theta)t)\,\mathrm{d}\theta+\int_0^\pi \delta(x+y+v_g(\theta)t)\,\mathrm{d}\theta\Bigg] \mathrm{d}y.
    \end{array}}
\end{equation}
\normalsize
It is shown from \edited{the solution}~(\ref{Slow1}) that  $T^S$ is represented as superposition of localized wave packets, \edited{which propagate} with the different group velocities \edited{and do not interact with each other}. This is the property of ballistic heat transport~(see, e.g.,~\cite{LISHI}). In contrast to the infinite chain, the wave packets can propagate both from the boundary~(is described by \edited{second} and third terms in~(\ref{Slow1})) and towards boundary~(is described by \edited{first} term in~(\ref{Slow1})). \edited{The} fourth term is equal to zero, because~$T^0(-x)=0$. Using the property of convolutions, we simplify the \edited{Eq.}~(\ref{Slow1}):
\begin{equation}\label{Slow2}
    T^S={\frac{1}{2\pi}\int_0^\pi T^0\,(\vert x+v_st\cos{\theta}\vert)\mathrm{d}\theta}.
\end{equation}

\small
\textit{Remark} 1. In~\cite{Kuz2021}, the heat transport in the infinite Hooke chain is investigated in the frameworks of both the lattice dynamics approach and the Boltzmann kinetic theory. Following the \edited{latter}, solution for the continuum kinetic temperature is derived using the distribution function as solution of the collisionless Boltzmann transport equation. The result coincides with predictions from the lattice dynamics approach. As for the semi-infinite free end Hooke chain, the kinetic temperature can be obtained in the same way, if the solution of the collisionless Boltzmann transport equation with evenness condition at the boundary~($x=0$) is known. This approach leads to the same result~(\edited{Eq.}~(\ref{Slow2})).
\normalsize
\subsection{Fast term}\label{subsect32}

Analogously, using the assumption~(\ref{LimREP}) with~$a\Delta N \ll \mathcal{A}$ and~$\Delta N \gg 1$ and~(\ref{LimREP2}), we obtain the  expression for the fast term~$T^F$. 
 Since the main contribution to the terms~$T^F$ comes from points~$\theta_1\approx \theta_2$, as it was shown is Sect.\ref{subsect31}, then $\omega(\theta_1)+\omega(\theta_2)\approx 2\omega(\theta_1)$.
 According to Sect.~\ref{subsect31}, the fundamental solution, $g_c^F$, can be written as
 \small
 \begin{equation}\label{ContF}
     g_c^F(x,y)=\frac{\delta(x-y)+\delta(x+y)}{2\pi}\int_0^\pi \cos{(2\omega(\theta)t)}\mathrm{d}\theta=\frac{\delta(x-y)+\delta(x+y)}{2}J_0(4\omega_e t).
 \end{equation}
 \normalsize
  Then, \edited{Eq.} for~$T^F$ has the form
 \begin{equation}\label{Fast2}
    T^F=\frac{T^0(x)H(x)+T^0(-x)H(-x)}{2}J_0(4\omega_e t)=\frac{T^0(x)}{2}J_0(4\omega_e t).
 \end{equation}
 Therefore, the expression for the fast term of the kinetic temperature, $T^F$, coincides with the same \edited{expression} for the infinite chain~\cite{Krivtsov2014}.

   Thus, the final expression for kinetic temperature in the continuum limit~(\ref{ContT}) is the sum of \edited{the contributions ~$T^F$, corresponding to the fast processes caused by equilibration of \edited{the} kinetic and potential energies and ~$T^S$,} corresponding to the slow processes caused by ballistic heat transport.

\section{Fundamental continuum solution}\label{sect4}

    We consider instantaneous thermal perturbation \edited{at some} point~$ha,\, h\in \mathbb{N} \cup \{0\}$. \edited{The initial temperature profile, corresponding to the considered case, is}
\begin{equation}\label{Deltapert}
    T^0(x)=A\delta(x-ha),
\end{equation}
where~$A$ is a constant with dimension~\edited{$\mathrm{K}\cdot \mathrm{m}$}. We write \edited{the continuum solution for the kinetic temperature} as\footnote{ \edited{The type of the initial temperature perturbation contradicts with the assumption, made in Sect.\ref{subsect30}. However, as it is shown below, the continuum solution has the same physical meaning, which is characteristic for one at arbitrary initial temperature profile.}}
\begin{equation}\label{InstPert}
    T\approx T^S=\frac{A}{2\pi}\int_0^\pi \delta(\vert x+v_s\cos{\theta}t \vert-ha)\mathrm{d}\theta.
\end{equation}
Here, we omit the term~$T^F$ because time scale of the fast process is much less than time scale of the ballistic heat transport.
Calculation of \edited{the integral}~(\ref{InstPert}) is carried out using the \edited{identity}~\cite{Gelfand}:
\begin{equation}\label{Gelf}
    \int_\mathcal{D} \delta(f(\xi))\mathrm{d}\xi=\sum_i \vert f'(\xi_i)\vert ^{-1},\qquad f(\xi_i)=0,
\end{equation}
where~$\xi_i$ are zeros of function~$f$, lying inside the domain~$\mathcal{D}$. Therefore, we have the following \edited{solution} for the kinetic temperature:
\begin{equation}\label{FundSol}
  T(x,t)=\frac{A}{2\pi}\left(\frac{H(v_st-\vert x-ha \vert)}{\sqrt{v_s^2t^2-(x-ha)^2}}+\frac{H(v_st-\vert x+ha \vert)}{\sqrt{v_s^2t^2-(x+ha)^2}}\right).
\end{equation}
We have obtained the continuum \edited{solution for kinetic temperature} in the semi-infinite chain in the case of instantaneous point heat pulse. However, if we consider the infinite chain with \edited{heat pulses} at the points~$ha$ and~$-ha$, the obtained continuum solution is the same~\edited{(it follows from Eq. for~$T^S$ in the infinite Hooke chain, see~(\ref{CONT_SOL}))}. Therefore, the continuum kinetic temperature field in the semi-infinite Hooke chain with free end and some source~\textit{coincides} with the \edited{one} in the infinite Hooke chain with the same and mirrored sources. The aforesaid rule will be further referred to as a~\textit{principle of continuum solution symmetry}. 
Thus, expression for kinetic temperature is represented as sum of two contributions. The first contribution is the solution for the infinite chain~\cite{Sokolov2021,Allen} and corresponds to the propagating incident waves. The second term corresponds to the wave, reflected from the free boundary. Note that, at times~$t<ha/v_s$, heat propagation can be described via the first term in~Eq.(\ref{FundSol}).

\begin{figure}[htb]
\center{\includegraphics[width=0.55\linewidth]{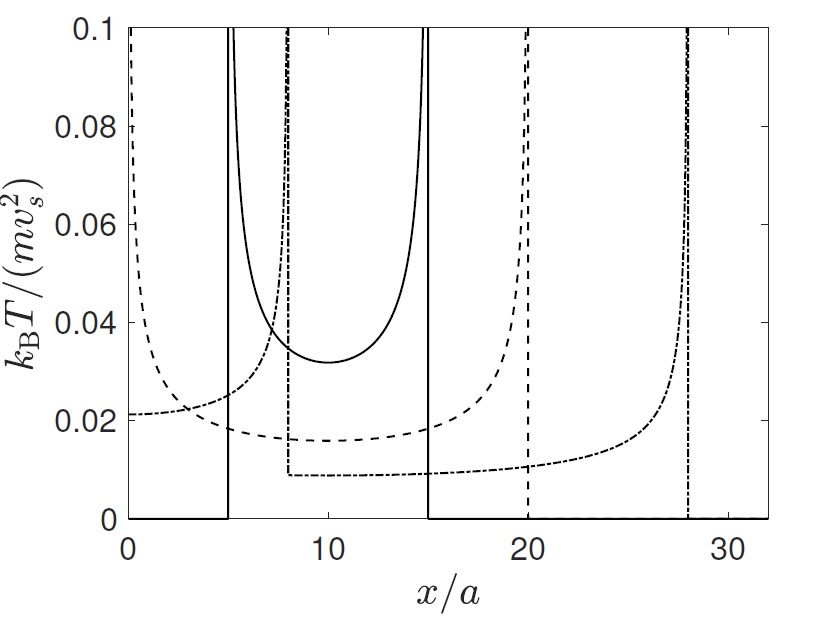}}
\caption{Thermal waves \edited{at} $\omega_e t=5$~(solid line), $\omega_e t=10$~(dashed line), $\omega_e t=18$~(dash-dotted line).}
\label{fig2}
\end{figure}

    We consider \edited{the} heat perturbation, located at some point from the boundary~($h=10$). Behavior of thermal waves, propagation of which obeys \edited{Eq.}~(\ref{FundSol}), is presented in Fig.~\ref{fig2}.
\edited{It is shown in Fig.\ref{fig2}} that the waves in the semi-infinite chain travel in both directions before and after reflection from the boundary. \edited{The} front instantly changes direction of propagation \edited{after the reflection}.  Specifically, it is shown that, the temperature profile is \textit{not symmetric} with respect to the heat source after the reflection.

    Thus, we have analytically described a property of the ballistically propagating thermal waves to reflect from \edited{free} boundaries. The analytical \edited{solution}, describing propagation of the waves before and after the reflection, is derived. In the next section, we compare \edited{the} continuum and discrete descriptions of the kinetic temperature field in cases of perturbation on the finite domain.

\section{Comparison of the discrete and continuum solutions}\label{sect5}
In this section, evolution of the kinetic temperature fields in the cases of rectangular and step thermal perturbations \edited{is} under \edited{consideration}. \edited{The initial temperature profiles are chosen for convenience of determination of the characteristic for heat propagation time scales, corresponding to the time for thermal wave to reach the boundary and to reflect from the boundary.} We compare \edited{the} discrete and continuum solutions for kinetic temperature fields and show that \edited{the} continuum description of heat transport \edited{has} some restrictions. 

In numerical simulations, the kinetic temperature is calculated by \edited{its definition}~(\ref{DefT}), where the mathematical expectation is replaced by average over~$R$ realizations. To obtain the particle velocity, we solve \edited{dynamical} equations~(\ref{DYNeq}) for the finite~\footnote{Simulations are performed for the chain with~$500$ particles.} chain with two free ends with initial conditions~\footnote{Random numbers~$\rho_n$ are uniformly distributed in the segment~$[-\sqrt{3}; \sqrt{3}]$, \edited{which satisfies the condition~(\ref{UNCS})}.}~(\ref{IC}) and~(\ref{defT0}) using the fourth-order Candy and Rozmus~\cite{Candy} integrator with the optimizing parameters, proposed in~\cite{Lach1992} and time step~$\Delta t$. The following parameters are used:
\begin{equation}\label{PARAM}
    R=10^5,\qquad \Delta t=0.01/\omega_e,
\end{equation}
where~$\omega_e$ is defined in~(\ref{DYNeq}).

 \subsection{Rectangular initial perturbation}\label{subsect52}
\begin{figure}[htb]
\begin{minipage}[H]{0.5\linewidth}
\center{\includegraphics[width=0.95\linewidth]{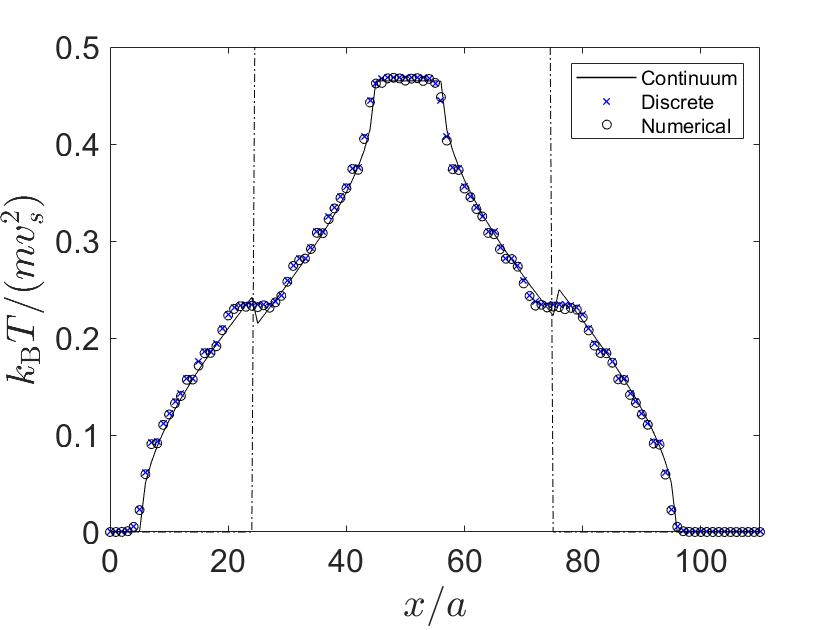}}
\end{minipage}
\begin{minipage}[H]{0.5\linewidth}
\center{\includegraphics[width=0.95\linewidth]{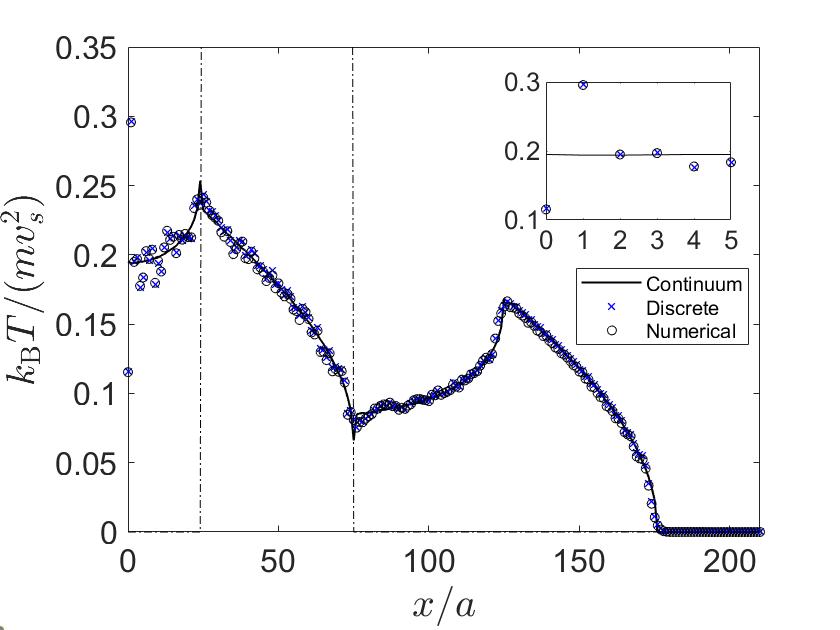}}
\end{minipage}
\caption{Discrete and continuum solutions for \edited{the} kinetic temperature in the semi-infinite \edited{free end} Hooke chain in the case of rectangular initial perturbation at $\omega_e t=20$~(left) and $\omega_e t=100$~(right). Width of \edited{the} initial thermal perturbation~(\ref{RECT}) is limited by the dash-dotted lines. }
\label{fig3}
\end{figure}

  Consider a rectangular heat perturbation, which is defined \edited{as}

\begin{equation}\label{RECT}
    k_\mathrm{B}T^0(x)=m v_s^2 \left(H(x-L_1)-H(x-L_1-L_2) \right),
\end{equation}
where~$L_1$ is a distance from the boundary to the perturbation; ~$L_2$ is a width of perturbation. We take~$L_1=25a$ and~$L_2=50a$ and \edited{the} investigate temperature profiles in two cases: before reflection of thermal wave and after \edited{the} reflection. Discrete and continuum solutions for kinetic temperature are \edited{presented} in Fig.~\ref{fig3}.
It is seen in Fig.~\ref{fig3}~(left) that the continuum and discrete solutions practically coincide before thermal wave reaches the boundary, except regions near wavefront and near the boundaries of initial perturbation. The mismatches of the \edited{discrete and continuum solutions} are caused by finiteness of the perturbation length and fast process, which takes place at relatively short times. The discrepancies, mentioned above, become infinitesimal at~$\omega_e t=100$~(see \edited{the} right Fig.~\ref{fig3}), when energy of the fast process is much less than \edited{the} energy transferred along the chain. However, the discrete \edited{solution} undergoes a jump near the boundary~(see inlet Fig.~\ref{fig3}). Therefore,  the discrete and continuum solutions~\textit{disagree} after reflection of thermal wave from the boundary. In order to investigate this jump in detail, we consider behavior of the \edited{kinetic} temperature at the boundary.

\begin{figure}[htb]
\center{\includegraphics[width=0.55\linewidth]{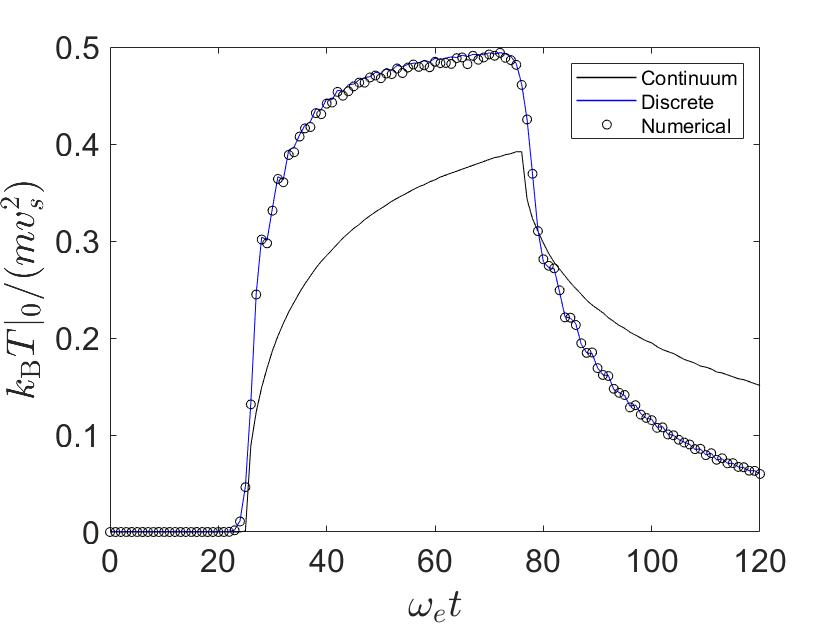}}
\caption{Evolution of the kinetic temperature at the boundary.}
\label{fig4}
\end{figure}

The \edited{continuum solution} for the \edited{kinetic} temperature at the boundary, $T(0,t)$, has the following form, which can be \edited{obtained} by substitution of~(\ref{RECT}) to~(\ref{ContT}) with~$x=0$ and subsequent integration from~$0$ to $\pi$:
\small
\begin{equation}\label{RECT_READY}
  k_\mathrm{B} T(0,t)=\frac{mv_s^2}{\pi}\Bigg[\arccos{\left(\frac{L_1}{v_st}\right)}H\left(t-\frac{L_1}{v_s}\right)-\arccos{\left(\frac{L_1+L_2}{v_st}\right)}H\left(t-\frac{L_1+L_2}{v_s}\right)\Bigg].
\end{equation}
\normalsize
    From~(\ref{RECT_READY}), one can conclude that evolution of the kinetic temperature at the boundary, caused by rectangular perturbation in the chain, has three stages. The first stage is related with fast processes and propagation of thermal waves before reflection~($t<L_1/v_s)$. The second stage, related to reflection of thermal wave, begins at~$t=L_1/v_s$ and has duration~$t=L_2/v_s$. Finally, the third stage begins after reflection of the wave from boundary at~$t=(L_1+L_2)/v_s$ and is related to relaxation of the temperature at the boundary, which decays~(according to the continuum solution) as~$1/t$. Evolution of the discrete and continuum solutions \edited{at the boundary} is presented in Fig.~\ref{fig4}.
One can see from the Fig.~\ref{fig4} that the discrete solution for the kinetic temperature at the boundary significantly differs from the continuum \edited{one} after front reaches the boundary. Growth of the discrete solution at the boundary, caused by reflection of thermal wave and decrease of this, caused by propagation of wavefront backwards, are faster than the same stages of evolution of the continuum solution. 

Thus, \edited{the} process of \edited{heat} transport in the semi-infinite Hooke chain caused by rectangular perturbation can be generally described by the continuum model, if we deal with the heat propagation far from the boundary. However, there are discrepancies between the continuum and discrete solutions near and at the boundary. 

 \subsection{Step initial perturbation}\label{subsect53}
\begin{figure}[htb]
\begin{minipage}[H]{0.5\linewidth}
\center{\includegraphics[width=0.95\linewidth]{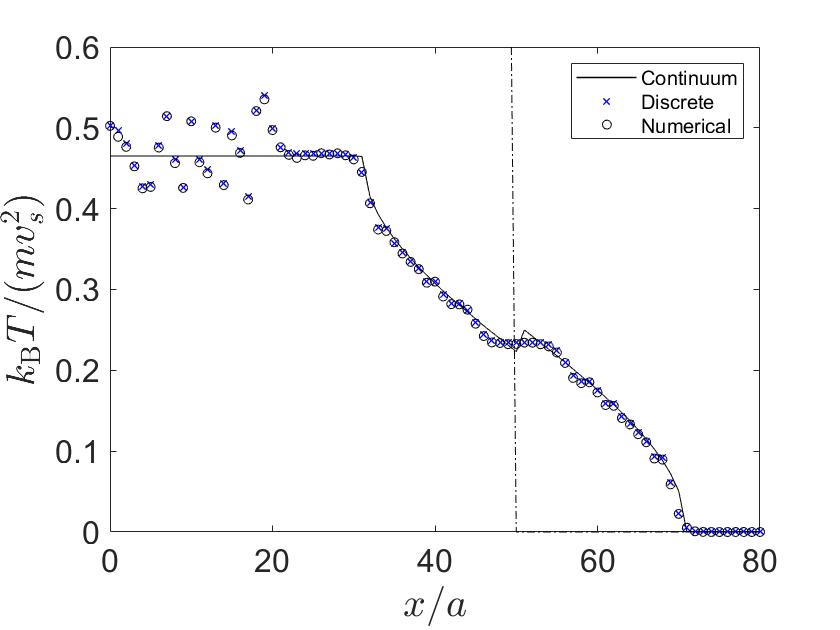}}
\end{minipage}
\begin{minipage}[H]{0.5\linewidth}
\center{\includegraphics[width=0.95\linewidth]{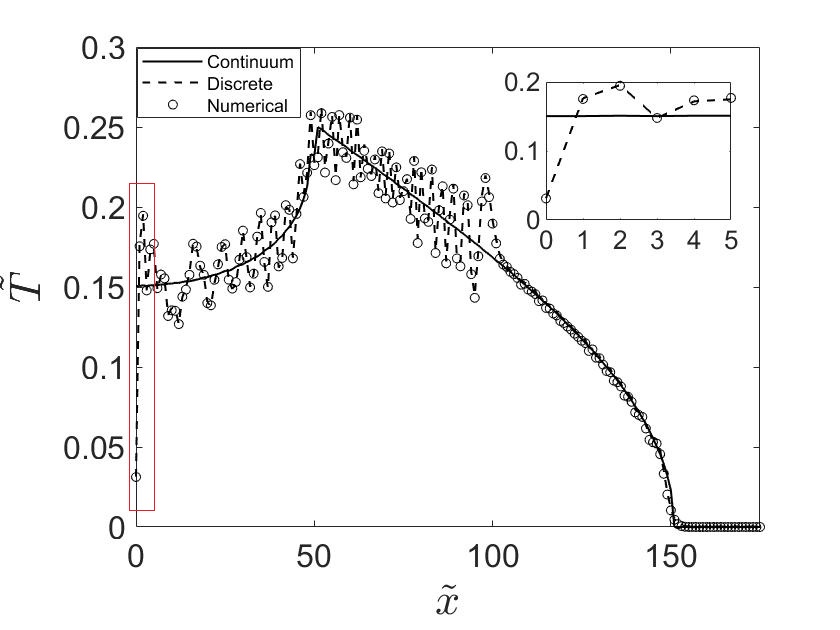}}
\end{minipage}
\caption{Discrete and continuum \edited{solutions} for the kinetic temperature in the semi-infinite \edited{free end Hooke} chain \edited{in the case of step initial perturbation at $\omega_e t=20$~(left) and $\omega_e t=100$~(right)}. Width of \edited{the} initial thermal perturbation~(\ref{STEP}) is limited by the dash-dotted line. }
\label{fig5}
\end{figure}

 Consider \edited{a} step heat perturbation:
 \begin{equation}\label{STEP}
   k_\mathrm{B} T^0(x)=mv_s^2H(L-x),
\end{equation}
where~$L$ is a width of perturbation. The discrete and continuum solutions, corresponding to the case, are presented in Fig.~\ref{fig5} for~$L=50a$.

It is shown in Fig.\ref{fig5} that \edited{the} discrete and continuum solutions are also significantly different near the boundary after reflection of thermal wave by virtue of the jump~(see inlet Fig.\ref{fig5}). Moreover, some mismatches between \edited{the} discrete and continuum solutions are observed both before~($t<L/v_s$) and after~($t>L/v_s$) reflection of thermal wave from the boundary. \edited{On the one hand}, these mismatches could be caused by influence of the fast process, occurring at the same time of the wavefront propagation~(see \edited{the} left Fig.\ref{fig5}). \edited{On the other hand}, however, it is seen from \edited{the} right Fig.\ref{fig5} that the differences between the discrete and continuum solutions remain after the reflection.  Indeed, one can see a perturbation, propagating along the chain approximately with the \edited{speed of sound}, which is described not by the continuum but rather by \edited{the} discrete solution.  An explanation of physical reason of this perturbation is beyond the scope of present paper.

\begin{figure}[htb]
\center{\includegraphics[width=0.55\linewidth]{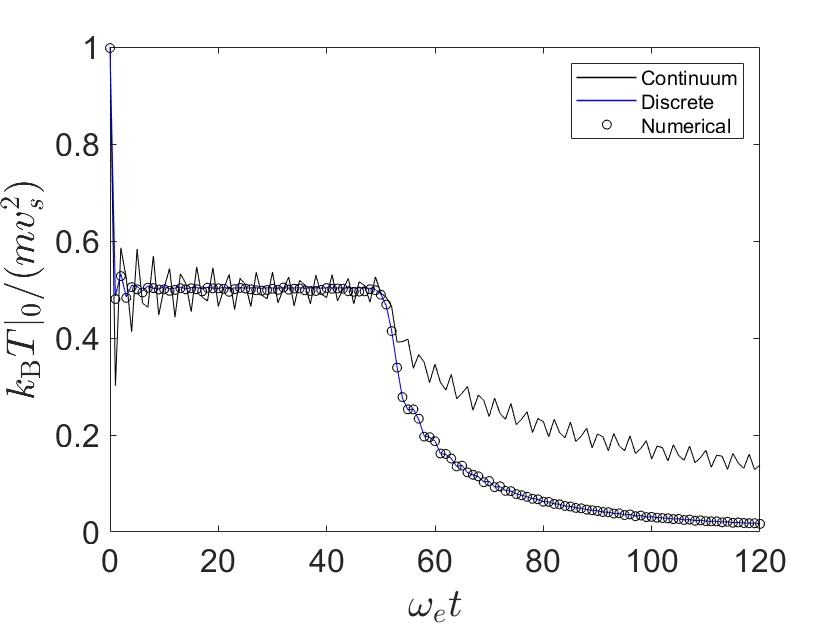}}
\caption{Evolution of the kinetic temperature at the boundary.}
\label{fig6}
\end{figure}

The continuum solution for \edited{the kinetic temperature} at the boundary can be expressed as
\small
\begin{equation}\label{STEP_READY}
  k_\mathrm{B}T(0,t)=mv_s^2\left(\frac{1}{2}\left(H\left(\DS \frac{L}{v_s}-t\right)+J_0(4\omega_e t)\right)+\frac{1}{\pi}\arcsin{\left(\frac{L}{v_st}\right)}H\left(t-\frac{L}{v_s}\right)\right).
\end{equation}
\normalsize
\edited{The} evolution of the kinetic temperature occurs in two stages: equilibration of the kinetic temperature~(at times~$t<L/v_s$), \edited{accompanied by the reflection of thermal wave}. At times~($t>L/v_s$) the wavefront propagates after \edited{the} reflection. Then the kinetic temperature at the boundary decays.\footnote{As in the case of rectangular perturbation, the continuum solution decays also as~$1/t$.} Evolution of functions of the \edited{kinetic} temperatures at the boundary are presented in Fig.\ref{fig6}.
\edited{It is shown} in Fig.~\ref{fig6}  behavior of the discrete and continuum kinetic temperature fields is significantly different not only at~$t>L/v_s$~(when discrete solution also asymptotically deviates from the continuum) but also at~$t<L/v_s$. 
\small

\textit{Remark} 2.
According to the preliminary calculations, both discrete and continuum solutions for the kinetic temperature at the boundary are scale invariant with respect to $L$.~\footnote{i.e. the function $T(t)\Big \vert_0$ at width of the initial thermal perturbation~$L+\Delta L$ is equal to~$T\left(\frac{L+\Delta L}{L}t\right)\Big \vert_0$ at the width~$L$.}
\normalsize

Thus, the \edited{continuum} solution for kinetic temperature significantly differs from the \edited{discrete one}, which is shown by \edited{a} jump \edited{of the latter} near the boundary. The discrete solution at the boundary decays substantially faster than \edited{the} continuum \edited{one}. This observation requires \edited{a} detailed asymptotic analysis based on the stationary phase method~\cite{Erdely, Fedoryuk} and therefore needs a separate investigation, which remains beyond the scope of present paper. 

In the next section, we compare \edited{the} discrete and continuum descriptions of heat transport in the semi-infinite and infinite chains.

\section{Kinetic temperatures in the infinite chain}\label{sect6}

In the section, we investigate heat transport in the \textit{infinite} Hooke chain in the case of two symmetric with respect to zero initial thermal perturbations, i.e.
\begin{equation}\label{BIRECT}
   k_\mathrm{B} T^0(x)=mv_s^2 \Big(H(x-L_1)-H\left(x-L_1-L_2\right)+H\left(x+L_1+L_2\right)-H(x+L_1)\Big),
\end{equation}
where the values~$L_1$ and~$L_2$ are defined in Sect.\ref{subsect52}. The initial temperature profile~(\ref{BIRECT}) \edited{implies} two mirrored with respect to zero heat sources. 

\begin{figure}[htb]
\begin{minipage}[H]{0.5\linewidth}
\center{\includegraphics[width=0.95\linewidth]{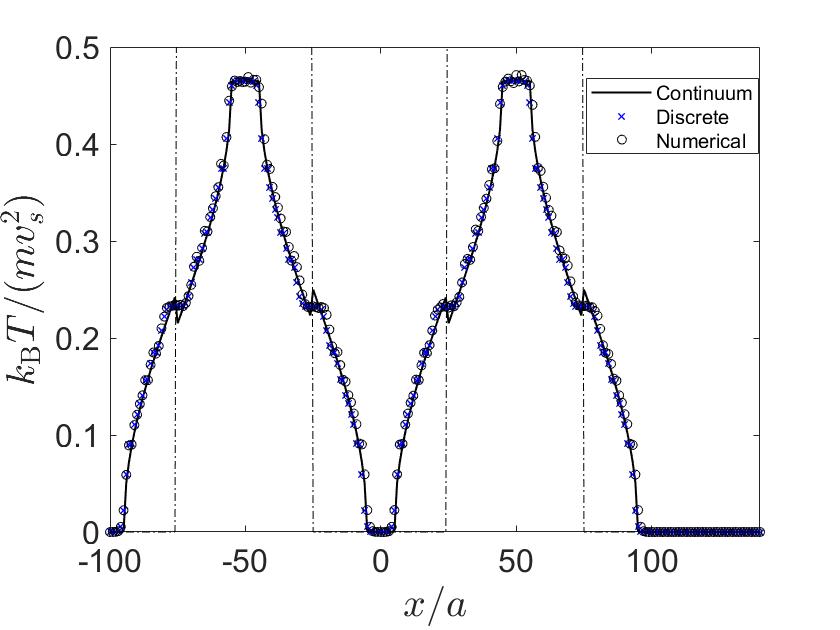}}
\end{minipage}
\begin{minipage}[H]{0.5\linewidth}
\center{\includegraphics[width=0.95\linewidth]{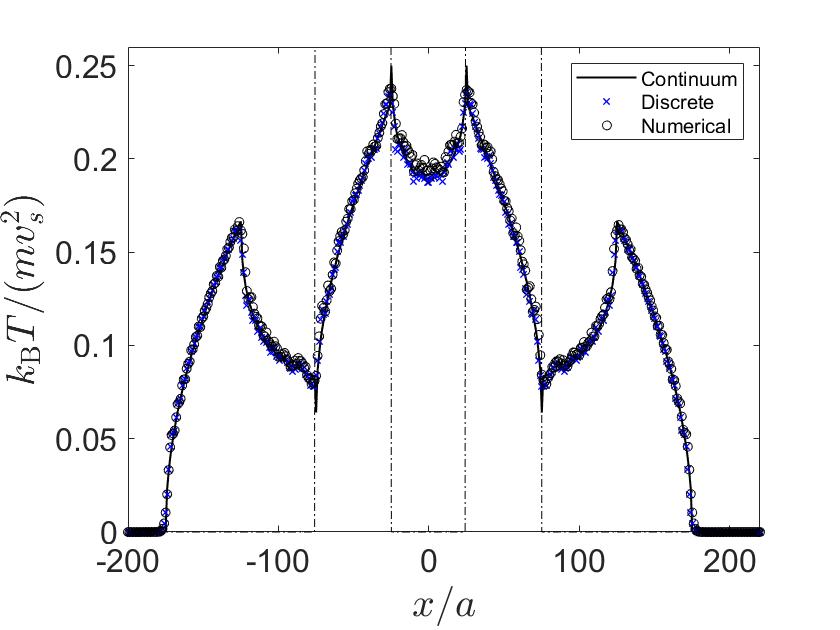}}
\end{minipage}
\caption{Discrete and continuum solutions for kinetic temperature in the infinite Hooke chain, in the case of \edited{double} rectangular initial perturbation~(\ref{BIRECT}) \edited{at} $\omega_et=20$~(left) and $\omega_et=100$~(right). Width of \edited{the} initial thermal perturbation~(\ref{BIRECT}) is limited by the dash-dotted lines. }
\label{fig8}
\end{figure}

Further, we consider \edited{the} two cases. The first case corresponds to symmetry of heat sources but uncorrelated initial velocities. Therefore, the governing \edited{dynamical} equations are

\begin{equation}\label{DLESS_DYN2}
    \ddot{u}_n=\omega_e^2(u_{n+1}-2u_n+u_{n-1}),\quad n\in \mathbb{Z} \setminus \{0\}.
\end{equation}
with initial conditions~(\ref{IC}), \edited{(\ref{defT0})} and~(\ref{UNCS}), corresponding to~(\ref{BIRECT}).
The problem is solved numerically in the same way as discussed in Sect.\ref{sect5} and \edited{the} periodic boundary conditions are used. Analytical solution of the problem is also presented both in the discrete~(see Eq.~(\ref{DisT})) and continuum descriptions. The continuum solution is proposed in~\cite{Kuz2017, Sokolov2021}:
\begin{equation}\label{CONT_SOL}
\begin{array}{l}
   \DS  T_\mathrm{inf}(x,t)=T^F_\mathrm{inf}(x,t)+T^S_\mathrm{inf}(x,t),\\[2mm]
    \DS T^F_\mathrm{inf}=\frac{T^0(x)}{2}J_0(4\omega_et),\quad \DS T^S_\mathrm{inf}=\frac{1}{2\pi}\int_{0}^\pi T^0(x+v_st\cos{\theta})\mathrm{d}\theta.
    \end{array}
\end{equation}
The discrete and continuum solutions are presented in Fig.~\ref{fig8} \edited{for the} different moments of time. 
It is seen from Fig.\ref{fig8} that the continuum and discrete solutions are in good agreement. Moreover, in the domain~$x\geqslant 0$ the continuum kinetic temperature field coincides with the same field in the semi-infinite chain\footnote{Therefore, before reflection from the boundary, the continuum solution for the kinetic temperature in the semi-infinite chain obeys \edited{the solution}~(\ref{CONT_SOL}).}. As expected, the principle of continuum solution, formulated in Sect.\ref{sect4}, is fulfilled. However, the corresponding discrete solutions disagree~(see \edited{the} right Fig.\ref{fig3}). The reason of the mismatch is the random initial velocities are uncorrelated. If so, then the discrete solutions are not symmetric with respect to zero even in case of mirrored heat sources. 

\begin{figure}[htb]
\center{\includegraphics[width=0.6\linewidth]{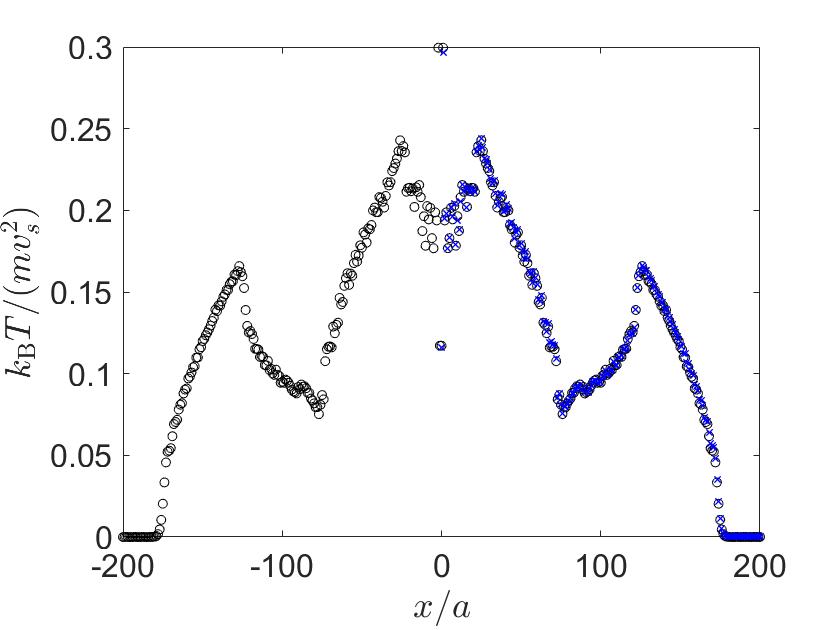}}
\caption{The discrete \edited{(numerical)} solution for the kinetic temperature \edited{field} in the infinite Hooke chain in the case of \edited{double} rectangular initial perturbation~(\ref{BIRECT}) with \edited{the} condition~(\ref{SYMMETR})~(black circles) and the \edited{discrete solution} for the kinetic temperature in the semi-infinite free end Hooke chain in the case of rectangular initial perturbation~(\ref{RECT})~\edited{(blue crosses)}. }
\label{fig9}
\end{figure}

Consider another case, which is governed by the \edited{dynamical} equations~(\ref{DLESS_DYN2}) with initial conditions~(\ref{IC}). In addition, we require the following condition for random numbers~$\rho_n$:
\begin{equation}\label{SYMMETR}
    \rho_{-n}=\rho_n,
\end{equation}
which implies a symmetry with respect to zero of both initial temperature profile and field of initial velocities simultaneously. 
To the best of our knowledge, analytical solution for the kinetic temperature~(both in the discrete and continuum formulations), corresponding to the problem~(\ref{DLESS_DYN2}) with~(\ref{SYMMETR}) is unknown. Therefore, we calculate numerically the kinetic temperature in the same way as discussed earlier~(for the case of uncorrelated initial velocities). Comparison between kinetic temperature field in the considered model at~$\omega_et=100$ with the corresponding solution for the semi-infinite chain~(\edited{Eqs.}~(\ref{ContT}) and~(\ref{DisT2})) is presented in Fig.\ref{fig9}.
 It is seen from~Fig.\ref{fig9} that the discrete solution for the semi-infinite chain in case of the rectangular perturbation is the same as solution for the infinite Hooke chain with exactly mirrored heat sources~(with symmetric initial velocity fields with respect to zero). Therefore, the discrete \edited{solution for} kinetic temperature in the semi-infinite Hooke chain with free end and some field of initial velocities coincides with the \edited{one} in the infinite Hooke chain with the same and mirrored \edited{initial} velocity fields. 

Thus, symmetry of the initial temperature field~(with respect to zero) is enough for symmetry of the continuum solution, but the symmetry of the discrete solution should be in addition provided by the symmetry of the initial velocity field.

\section{Conclusions}\label{sect7}

In the paper, we have \edited{investigated the} process of unsteady heat transfer in the semi-infinite free end Hooke chain. We have proposed both discrete~(exact) and continuum descriptions of this process, studying evolution of the kinetic temperature after \edited{an} instantaneous heat pulse. 

The discrete kinetic temperature field can be constructed by discrete analogue of convolution of the initial temperature profile and \edited{the} corresponding \edited{time derivative of the} fundamental solution~(see \edited{Eqs.}~(\ref{conv}) and~(\ref{fundPhi})). On the other hand, the solution can be obtained by symmetry property of the kinetic temperature field in the infinite chain with exactly mirrored heat sources~(the condition~(\ref{SYMMETR}), implying symmetry of the initial velocity field, is fulfilled). In turn, the continuum solution for the semi-infinite chain coincides with the continuum solution for the infinite chain with mirrored heat sources, \edited{corresponding} to the initial kinetic temperature field.

    It was analytically shown that process of ballistic heat propagation in the chain has at least two transient processes. The first is associated with reflection of thermal wave from the free boundary. The second transient process is related with subsequent propagation of thermal waves backwards. \edited{Comparison of the discrete and continuum solutions for the kinetic temperature revealed mismatch between themselves near the boundary and at the boundary during the transient processes. Indeed, the continuum solution does not change in space near the boundary, where the discrete solution undergoes a jump.  Based on aforesaid, we conclude that the continuum description~(which may be obtained, e.g., from the Boltzmann kinetic theory) of ballistic heat transport in the lattices with boundary conditions needs clarification with taking into account discreteness of the lattices.} 
    
    \edited{The continuum solution can be improved at least in two ways.} The first is obtaining large time asymptotics for the discrete \edited{solution}~(either through \edited{Eq.}~(\ref{DisT2}) in the way, as proposed in~{\cite{Gavr2022}}, or through asymptotic solution of Eq.~(\ref{fundPhi}) in the way, as proposed in~\cite{Gavr_isot}) and then to proceed to the continuum limit. The second way is to solve heat transport problem in the infinite chain with mirrored with respect to zero fields of initial velocities and then perform a continualization procedure, as proposed either in~\cite{Kuz2021} or in~\cite{Gavr2018}. The improved continuum solution can be used as the constitutive relation~(in particular, for problems of thermoelasticity~(see~\cite{Kuz2020}) or thermoelectricity~(see, e.g.,~\cite{Ivanova2022})).
    
    An explanation of mechanism of origin of the kinetic temperature jump near the boundary \edited{has not been} provided by the model of the semi-infinite chain \edited{yet}. In order to understand this, heat transport through the boundary of the two chains with significantly different stiffnesses was numerically investigated, analogously as discussed in~Sect.\ref{sect5}. In this system, the jump of the kinetic temperature is the Kapitza jump, which was discovered experimentally long time ago~\cite{Kapitza}. From these observations, one can assume that the jump of the kinetic temperature near the free end is the limiting case of the Kapitza jump in the two interacting chains with~\edited{equal masses} and stiffnesses, ratio of which is infinitesimal. However, this assumption requires a confirmation based on the analytical treatments. The problem, considered and solved in the present paper, can be auxiliary. In general, a problem of heat transport in the heterogeneous lattices is as yet hard to solve analytically but some progress in studying it is attained in both the steady-state~\edited{\cite{Gendelman1, Gendelman2, Tian1}} and non-stationary~\edited{\cite{Gavr_isot, Tian2}} formulations.
    
    The results of the present paper \edited{are expected to be important for comprehensive understanding of unsteady thermal processes in the lattices with free boundaries and thus to} development of full-fledged theory of heat transport \edited{therein, which can be verified} by the experiments, \edited{described, e.g., in~\cite{Anufriev2, Wolfe, Northrop, Ravi, Qian}}. However, real systems are generally anharmonic~(nonlinear) and therefore investigation of heat propagation therein should take into account nonlinearity. On the other hand, heat transport regime remains quasiballistic in weakly anharmonic lattices at relatively short times and can be therefore qualitatively described in the harmonic approximation~(see, e.g.,~\cite{Kuz2020, Gendelman1, Korzn2020, MiM}). In particular, it is shown in paper~\cite{Gendelman1} that the jump of the temperature in the neighborhood of isotopic defect is preserved for insufficient time. However, the jump disappears in process of long time owing to nonlinearity. We assume that, \edited{depending on width of the initial heat pulse and on distance of the latter from the boundary,} the similar effect may be likely observed in the nonlinear semi-infinite lattices. 
    
    \section{Acknowledgements}\label{sect8}

The work is supported by the Russian Science Foundation~(Grant No.\;21-71-10129). The author is deeply grateful to V.A. Kuzkin, A.M. Krivtsov, S.N. Gavrilov, E.V. Shishkina, A.A. Sokolov, A.S. Murachev, \edited{N.M. Bessonov} and E.F. Grekova for useful and stimulating discussions \edited{and to anonymous referees for the valuable comments}. 

    
 \section*{A Derivation of the discrete analogue of fundamental solution}\label{A}
 Here, we derive the discrete fundamental solution, namely~$g_{n,j_s}^S(\Delta N)$. 
    Expanding a product of cosines in~(\ref{Divid}) yields 
\footnotesize
 \begin{equation}\label{eq45}
 \begin{array}{l}
    \DS S_{nj}=\frac{1}{16\pi^2}\iint_{-\pi}^\pi \Bigg[\cos ({(n+j+1)(\theta_1+\theta_2)})+ \cos ({(n+j+1)(\theta_1-\theta_2)})+ \\[3mm]
    \DS \cos({(n-j)(\theta_1+\theta_2)})+\cos({(n-j)(\theta_1-\theta_2)})+\DS\cos{\left(\frac{(2j+1)(\theta_1+\theta_2)}{2}-\frac{(2n+1)(\theta_1-\theta_2)}{2}\right)}+\\[3mm]
    \DS\cos{\left(\frac{(2j+1)(\theta_1+\theta_2)}{2}+\frac{(2n+1)(\theta_1-\theta_2)}{2}\right)}+\cos{\left(\frac{(2n+1)(\theta_1+\theta_2)}{2}-\frac{(2j+1)(\theta_1-\theta_2)}{2}\right)}+\\[3mm]
    \DS\cos{\left(\frac{(2n+1)(\theta_1+\theta_2)}{2}+\frac{(2j+1)(\theta_1-\theta_2)}{2}\right)}\Bigg] \cos \left({(\omega(\theta_1)-\omega(\theta_2)) t}\right) \mathrm{d}\theta_1 \mathrm{d}\theta_2.
\end{array}
 \end{equation}
\normalsize
Therefore, the expression for~$g_{n,j_s}^S(\Delta N)$ can be rewritten as a sum of the following eight terms:
\small
  \begin{equation}\label{eq46}
 \begin{array}{l}
 \DS g_{n,j_s}^S(\Delta N)=\frac{1}{16\pi^2 a}\iint_{-\pi}^\pi \cos{\left((\omega(\theta_1)-\omega(\theta_2)) t\right)}\sum_{i=1}^8 \varphi_i(\theta_1,\theta_2) \mathrm{d}\theta_1 \mathrm{d}\theta_2,\\[2mm]
 \DS \varphi_1(\theta_1,\theta_2)=\frac{1}{2\Delta N}\sum_{j=j_s-\Delta N+1}^{j_s+\Delta N}\cos{((n+j+1)(\theta_1+\theta_2))}=\\[4.5mm]
 \DS\frac{1}{2\Delta N}\cos{\Bigl(\frac{3(\theta_1+\theta_2)}{2}+(\theta_1+\theta_2)(n+j_s)\Bigr)}\frac{\sin {\left ((\theta_1+\theta_2)\Delta N \right )}}{\sin{\left(\frac{\theta_1+\theta_2}{2}\right)}},\\[4mm]
 \DS \varphi_2(\theta_1,\theta_2)=\frac{1}{2\Delta N}\sum_{j=j_s-\Delta N+1}^{j_s+\Delta N}\cos{((n+j+1)\Delta \theta)}=\\[4mm]
 \DS\frac{1}{2\Delta N}\cos{\left(\left (n+j_s+\frac{3}{2}\right)\Delta \theta\right)}\frac{\sin{(\Delta N \Delta \theta)}}{\sin{\frac{\Delta \theta}{2}}},\\[4mm]
  \DS \varphi_3(\theta_1,\theta_2)=\frac{1}{2\Delta N}\sum_{j=j_s-\Delta N+1}^{j_s+\Delta N}\cos{((n-j)(\theta_1+\theta_2))}=\\[4mm]\DS \frac{1}{2\Delta N}\cos{\left((\theta_1+\theta_2)\left(\frac{1}{2}+j_s-n\right)\right)}
  \DS \frac{\sin {\left ((\theta_1+\theta_2)\Delta N \right )}}{\sin{\left(\frac{\theta_1+\theta_2}{2}\right)}},\\[3mm]
  
 \DS \varphi_4(\theta_1,\theta_2)=\frac{1}{2\Delta N}\sum_{j=j_s-\Delta N+1}^{j_s+\Delta N}\cos{((n-j)\Delta \theta)}=\\[4mm]
 \DS \frac{1}{2\Delta N}\cos{\left(\Delta\theta\left(\frac{1}{2}+j_s-n\right)\right)}\frac{\sin{(\Delta N \Delta \theta)}}{\sin{\frac{\Delta \theta}{2}}},
 \end{array}
 \end{equation}
 \begin{equation*}
 \begin{array}{l}
 \DS\varphi_5(\theta_1,\theta_2)=\frac{1}{2\Delta N}\sum_{j=j_s-\Delta N+1}^{j_s+\Delta N}\cos{\left(\frac{(2j+1)(\theta_1+\theta_2)}{2}-\frac{(2n+1)(\theta_1-\theta_2)}{2}\right)}=\\[4mm]
 \DS=\frac{1}{2\Delta N}\cos{\left(\theta_1\left(\frac{1}{2}+j_s-n\right)+\theta_2\left(\frac{3}{2}+j_s+n\right)\right)} \frac{\sin {\left ((\theta_1+\theta_2)\Delta N \right )}}{\sin{\left(\frac{\theta_1+\theta_2}{2}\right)}},\\[4mm]
\DS\varphi_6(\theta_1,\theta_2)=\frac{1}{2\Delta N}\sum_{j=j_s-\Delta N+1}^{j_s+\Delta N}\cos{\left(\frac{(2j+1)(\theta_1+\theta_2)}{2}+\frac{(2n+1)(\theta_1-\theta_2)}{2}\right)}=\\[4mm]
 \DS=\frac{1}{2\Delta N}\cos{\left(\theta_2\left(\frac{1}{2}+j_s-n\right)+\theta_1\left(\frac{3}{2}+j_s+n\right)\right)} \frac{\sin {\left ((\theta_1+\theta_2)\Delta N \right )}}{\sin{\left(\frac{\theta_1+\theta_2}{2}\right)}},\\[4mm]
\DS\varphi_7(\theta_1,\theta_2)=\frac{1}{2\Delta N}\sum_{j=j_s-\Delta N+1}^{j_s+\Delta N}\cos{\left(\frac{(2n+1)(\theta_1+\theta_2)}{2}-\frac{(2j+1)(\theta_1-\theta_2)}{2}\right)}=\\[4mm]
 \DS=\frac{1}{2\Delta N}\cos{\left(\theta_1\left(\frac{1}{2}+j_s-n\right)-\theta_2\left(\frac{3}{2}+j_s+n\right)\right)} \frac{\sin {\left (\Delta N\Delta \theta \right )}}{\sin{\left(\frac{\Delta \theta}{2}\right)}},\\[4mm]
 \DS\varphi_8(\theta_1,\theta_2)=\frac{1}{2\Delta N}\sum_{j=j_s-\Delta N+1}^{j_s+\Delta N}\cos{\left(\frac{(2n+1)(\theta_1+\theta_2)}{2}+\frac{(2j+1)(\theta_1-\theta_2)}{2}\right)}=\\[4mm]
 \DS=\frac{1}{2\Delta N}\cos{\left(\theta_2\left(\frac{1}{2}+j_s-n\right)-\theta_1\left(\frac{3}{2}+j_s+n\right)\right)} \frac{\sin {\left (\Delta N\Delta \theta \right )}}{\sin{\left(\frac{\Delta \theta}{2}\right)}}, \quad \Delta \theta=\theta_1-\theta_2.
     \end{array}
 \end{equation*}
 \normalsize
We rewrite the components~$\varphi_2, \varphi_4, \varphi_7, \varphi_8$, containing the difference of wave numbers~$\Delta \theta$ as follows:
\small
 \begin{equation}\label{eq47}
 \begin{array}{l}
 \DS\varphi_2=\frac{\Delta{\theta}}{2}\Bigg[\frac{\cos{\frac{3\Delta \theta}{2}}}{\sin{\frac{\Delta \theta}{2}}}\left(\cos{\left((n+j_s)\Delta \theta\right)}\right)-\frac{\sin{\frac{3\Delta \theta}{2}}}{\sin{\frac{\Delta \theta}{2}}}\left(\sin{\left((n+j_s)\Delta \theta\right)}\right)\Bigg]\mathrm{sinc}(\Delta N \Delta \theta),\\[2mm]
\normalsize
\DS  \mathrm{sinc}(x)=\frac{\sin{x}}{x},\\[4mm]
\small
\DS \varphi_4=\frac{\Delta \theta}{2}\Bigg[\cot{\frac{\Delta \theta}{2}}\cos{((n-j_s)\Delta \theta)}+\sin{((n-j_s)\Delta \theta)}\Bigg]\mathrm{sinc}(\Delta N \Delta \theta),\\[4mm]
\DS\varphi_7=\frac{\Delta \theta}{2}\Bigg[\frac{\cos{\frac{3\Delta \theta}{2}}}{\sin{\frac{\Delta \theta}{2}}} \cos{(\Delta \theta(n+j_s)-\theta_1(2n+1))}-\frac{\sin{\frac{3\Delta \theta}{2}}}{\sin{\frac{\Delta \theta}{2}}}\sin{(\Delta \theta(n+j_s)-\theta_1(2n+1))}\Bigg]\\[2mm]
\mathrm{sinc}(\Delta N \Delta \theta),\\[4mm]
\DS\varphi_8=\frac{\Delta \theta}{2}\Bigg[\cot{\frac{\Delta \theta}{2}} \cos{(\theta_1(2n+1)-\Delta \theta(n-j_s))}-\sin{(\theta_1(2n+1)-\Delta \theta(n-j_s))}\Bigg]\\[2mm]
\mathrm{sinc}(\Delta N \Delta \theta).
 \end{array}
 \end{equation}
 \normalsize
The Eq.~(\ref{eq47}) can be simplified due to our assumptions about continualization~\edited{(see Sect.\ref{subsect30})}. For $\Delta N \gg 1$, the function $\mathrm{sinc}(x)$ is equal to 1 if $\Delta \theta$ is zero and fast tends to zero if $\Delta \theta$ is not equal to zero. Therefore, the main contribution to the function~$g_{n,j_s}^S(\Delta N)$ comes from two close wavenumbers~$\theta_1, \theta_2$. Therefore, in the limit cases of~$\Delta \theta \rightarrow 0$ and $\Delta N~\gg 1$, we have
 \begin{equation}\label{eq48}
 \begin{array}{l}
  \DS  \varphi_2=\cos{((n+j_s)\Delta \theta)}\mathrm{sinc}(\Delta N \Delta \theta)+O\left(\frac{1}{\Delta N}\right),\\[2mm]
    \DS \varphi_4=\cos{((n-j_s)\Delta \theta)}\mathrm{sinc}(\Delta N \Delta \theta)+O\left(\frac{1}{\Delta N}\right),\\[2mm]
    \DS \varphi_7=\cos({\theta_1(2n+1)-(n+j_s)\Delta \theta})\mathrm{sinc}(\Delta N \Delta \theta)+O\left(\frac{1}{\Delta N}\right),\\[2mm]
    \DS \varphi_8=\cos({\theta_1(2n+1)-(n-j_s)\Delta \theta})\mathrm{sinc}(\Delta N \Delta \theta)+O\left(\frac{1}{\Delta N}\right),\\[2mm]
    \DS \varphi_1=\varphi_3=\varphi_5=\varphi_6=O\left(\frac{1}{\Delta N}\right).
     \end{array}
 \end{equation}
 The difference~$\omega(\theta_1)-\omega(\theta_2)$ can be decomposed into series:
 \begin{equation}\label{eq49}
 \omega(\theta_1)-\omega(\theta_2)\approx\omega'(\theta_1)\Delta \theta.
 \end{equation}
Substitution of~(\ref{eq48}), (\ref{eq49}) to~(\ref{eq46}) with dropping out of the terms of order~$O\left(\frac{1}{\Delta N}\right)$ gives the \edited{expression}~(\ref{DiscFUND}).

  \section*{B Derivation of \edited{expressions} for wave-packets in the limit of mesoscale}\label{B}
  
We show that approximation of expressions for wave packets in~(\ref{wpcgs}) in the limit case~($\Delta N \gg 1$) approaches us to the Fourier transform of the~$\mathrm{sinc}$ function.
Indeed,
\begin{equation}\label{eq50}
    \begin{array}{l}
    \DS\frac{1}{2\pi a}\int_{\theta-\pi}^{\theta+\pi}\cos{(q \Xi)}\mathrm{sinc}(q \Delta N)\mathrm{d}q=\frac{1}{2\pi a \Delta N}\int_{(\theta-\pi)\Delta N}^{(\theta+\pi)\Delta N}\cos{\left(\frac{q \Xi}{\Delta N}\right)}\mathrm{sinc}\,q \mathrm{d}q\\[2mm]
    \DS\approx \frac{1}{2\pi a \Delta N}\int_{-\infty}^{\infty}\cos{\left(\frac{q \Xi}{\Delta N}\right)}\mathrm{sinc}\,q \mathrm{d}q=\mathrm{Re}\left(\frac{1}{2\pi a \Delta N}\int_{-\infty}^{\infty}e^{\mathrm{i}\left(\frac{q \Xi}{\Delta N}\right)}\mathrm{sinc}\,q \mathrm{d}q\right).
    \end{array}
\end{equation}
Analogously,
\begin{equation}\label{eq51}
    \DS\frac{1}{2\pi a}\int_{\theta-\pi}^{\theta+\pi}\sin{(q \Xi)}\mathrm{sinc}(q \Delta N)\mathrm{d}q\approx \mathrm{Im}\left(\frac{1}{2\pi a \Delta N}\int_{-\infty}^{\infty}e^{\mathrm{i}\left(\frac{q \Xi}{\Delta N}\right)}\mathrm{sinc}\,q \mathrm{d}q\right).
\end{equation}
Since
    \begin{equation}\label{eq52}
        \frac{1}{2\pi}\int_{-\infty}^\infty e^{\mathrm{i}\xi q}\mathrm{sinc}\,q\;\mathrm{d}q=\frac{1}{2}H\left(1-\vert \xi \vert\right),
    \end{equation}
then one gets
\begin{equation}\label{eq53}
\begin{array}{l}
    \DS\frac{1}{2\pi a \Delta N}\int_{-\infty}^{\infty}\cos{\left(\frac{q \Xi}{\Delta N}\right)}\mathrm{sinc}\,q \;\mathrm{d}q=\frac{1}{2a\Delta N}H\left(1-\frac{ \vert \Xi \vert}{\Delta N}\right),\\[3mm]
    \DS\frac{1}{2\pi a \Delta N}\int_{-\infty}^{\infty}\sin{\left(\frac{q \Xi}{\Delta N}\right)}\mathrm{sinc}\,q \;\mathrm{d}q=0.
    \end{array}
\end{equation}

\end{document}